\documentclass[aps,pra,twocolumn,groupaddress]{revtex4}
\usepackage{graphicx}
\usepackage{amsmath,amssymb}
\usepackage{bm}
\usepackage{subfigure}
\usepackage{color}
\usepackage{bbold}
\begin{document}
\title{Schwinger-Keldysh theory for Bose-Einstein condensation of photons in a dye-filled optical microcavity}
\author{A.-W. de Leeuw}
\email{A.deLeeuw1@uu.nl}
\affiliation{Institute for Theoretical Physics, Utrecht
University, Leuvenlaan 4, 3584 CE Utrecht, The Netherlands}
\author{H.T.C. Stoof}
\affiliation{Institute for Theoretical Physics, Utrecht
University, Leuvenlaan 4, 3584 CE Utrecht, The Netherlands}
\author{R.A. Duine}
\affiliation{Institute for Theoretical Physics, Utrecht
University, Leuvenlaan 4, 3584 CE Utrecht, The Netherlands}
\date{\today}
\begin{abstract}
We consider Bose-Einstein condensation of photons in an optical cavity filled with dye molecules that are excited by laser light. By using the Schwinger-Keldysh formalism we derive a Langevin field equation that describes the dynamics of the photon gas, and in particular its equilibrium properties and relaxation towards equilibrium. Furthermore we show that the finite lifetime effects of the photons are captured in a single dimensionless damping parameter, that depends on the power of the external laser pumping the dye. Finally, as applications of our theory we determine spectral functions and collective modes of the photon gas in both the normal and the Bose-Einstein condensed phase.
\end{abstract}
\maketitle
\vskip2pc
\section{Introduction}
\label{sec:int}
After the theoretical prediction of Bose-Einstein condensation (BEC) in 1925 \cite{Bose,Einstein}, it took until 1995 for the first direct experimental observation of this phenomenon in weakly interacting atomic vapors \cite{BEC1,BEC2,BEC3}. In addition to these atomic gases, BEC of bosonic quasiparticles such as magnons \cite{BECmagnon}, exciton-polaritons \cite{BECpolariton,BECpolariton2} and photons \cite{BECphoton} is now also observed. The Bose-Einstein condensates of these quasiparticles form a different class of condensates as they are not in true thermal equilibrium. 
\newline
\indent These non-equilibrium Bose-Einstein condensates are driven by external pumping to compensate for the particle losses and thereby to keep the average number of particles in the system at a constant level. In these systems the steady state of the Bose gas is therefore determined by interparticle interactions that lead to quasi-equilibration and by the balance between pumping and particle losses. Furthermore, contrary to dilute atomic gases, the temperature is typically constant in these experiments. Instead, one varies the strength of the external pumping power while keeping the system at a constant temperature. Above some critical value of the pumping power, the density of particles in the system is above the critical density and the system undergoes BEC.
\newline
\indent Another special feature of these pumped systems is the temperature at which BEC occurs. Since BEC happens when the phase-space density is of the order of unity \cite{PS}, the temperature at which the magnons, exciton-polaritons and photons condense is inversely related to their mass to the power $3/2$. Although these particles do not even always have a bare mass, they are all formally equivalent to bosons with an effective mass that is several orders of magnitude smaller than that of alkali atoms. Therefore, these systems undergo BEC at temperatures in the range of 10 - 300 K instead of in the nK regime relevant for the atomic Bose-Einstein condensates. 
\newline
\indent In order to get a detailed understanding of these non-equilibrium Bose-Einstein condensates, we from now onwards focus on the photon experiment of Klaers \emph{et al.} \cite{BECphoton}. This experiment is concerned with a photon gas in a dye-filled optical resonator. The distance between the cavity mirrors is chosen such that the emission and absorption of photons with a certain momentum in the longitudional direction dominates over that of other momenta. Thus, this component of the momentum of the photons is fixed and the photon gas becomes equivalent to a Bose gas with a small effective mass. Furthermore, the gas becomes effectively two-dimensional. In general this prohibits observing BEC at non-zero temperature, since a
homogeneous two-dimensional Bose gas can only condense at zero temperature \cite{PS}. However, due to the curvature of the cavity mirrors there is a harmonic potential for the photons. Therefore, BEC of photons is observed above some critical pumping power, since a harmonically trapped two-dimensional Bose gas can exhibit BEC at a non-zero temperature \cite{BECharm1,BECharm2}.
\newline
\indent Theoretically, a lot of progress has been made for BEC of magnons and exciton-polaritons \cite{Magnon1,Magnon2,Polariton1,Polariton2,Polariton3,Polariton4,Polariton5}. Although the observation of Bose-Einstein condensation of photons is more recent, it has also motivated theoretical studies: Klaers \emph{et al.} predicted a regime of large fluctuations of the condensate number \cite{Theory1}. Furthermore, the authors of Ref. \cite{Theory2} found that the photons cannot reach thermal equilibrium for small absorption and emission rates. The modification of the Stark shift of an atom in a Bose-Einstein condensate of photons was investigated in Ref. \cite{Theory3}, and conditions for BEC of photons that are in thermal equilibrium with atoms of dilute gases were derived in Ref. \cite{Theory4}.
\newline
\indent In this article we develop a theory for the photon experiment performed by Klaers \emph{et al.} \cite{BECphoton}. In Sec. \ref{sec:model} we derive an effective action for the photons.  In Sec. \ref{sec:nonequi} we use this effective action to derive a Langevin field equation for the photons including Gaussian noise, that incorporates the effect of thermal and quantum fluctuations. The main advantage of this approach is that it simultaneously treat coherent and incoherent effects. In particular, it enables us to describe the complete time evolution of the photons including the relaxation towards equilibrium, thus also equilibrium properties can be obtained. Subsequently, we show that the finite lifetime effects of the photons can be captured in a single dimensionless parameter $\alpha$, that depends on the power of the external laser pumping the dye. In Sec. \ref{sec:equil} we calculate equilibrium properties of the homogeneous photon gas in the normal and Bose-Einstein condensed phase such as spectral functions, collective modes and damping. We end with conclusions and outlook in Sec. \ref{sec:concl}.

\begin{figure}[t]
 \centerline{\includegraphics{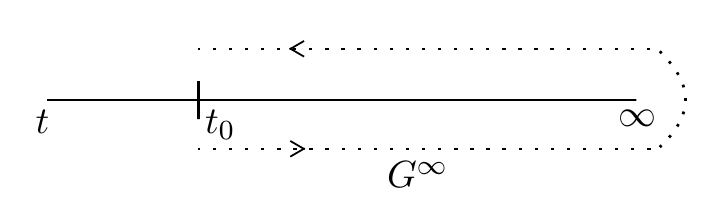}}
 \caption{The Schwinger-Keldysh contour $G^{\infty}$. The integration is first from $t_{0}$ to $\infty$ and then back from $\infty$ to $t_{0}$.}
 \label{fig:SK}
\end{figure}
\section{Model}
\label{sec:model}
\begin{figure}[t]
 \centerline{\includegraphics{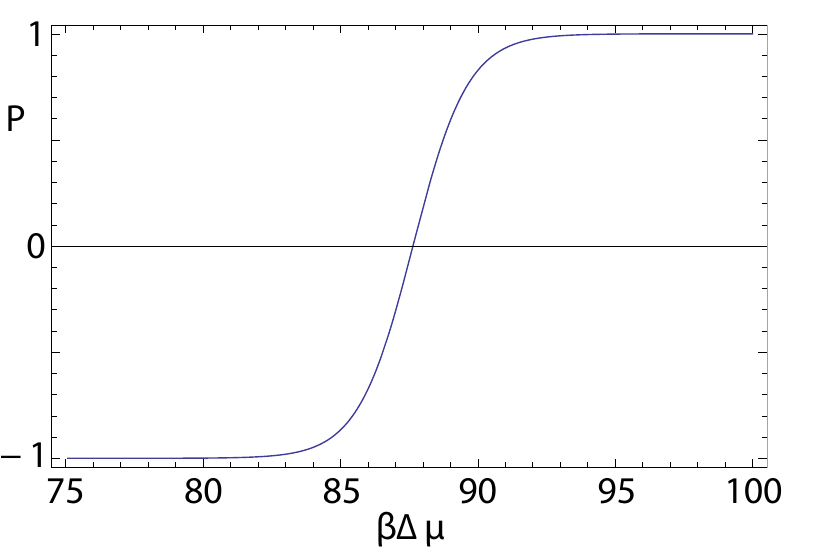}}
 \caption{A plot of the polarization of the molecules $P$ at room temperature $T = 300 \, \mathrm{K}$ as a function of $\beta\Delta\mu$ for a density of molecules $n_{\mathrm{m}} = 9 \cdot 10^{23} \, \mathrm{m}^{-3}$, $\Delta = 3.63 \cdot 10^{-19} \, \mathrm{J}$. The polarization is exactly zero if $\Delta\mu$ is equal to the energy difference between the excited and ground states of the molecules.}
 \label{fig:polarization}
\end{figure}

In this section we derive an effective action for the photons by using the Schwinger-Keldysh formalism developed in Ref. \cite{Stoofkeldysh}. In particular, the photons are coupled to a reservoir of dye molecules. The energy of the photons is given by 
\begin{align}\label{eq:enphot}
\epsilon_{\gamma}({\bf k}) = \hbar c \sqrt{k_{x}^{2} + k_{y}^{2} + k_{\gamma}^{2}},
\end{align}
where $\hbar$ is Planck's constant, $c$ is the speed of light in the medium and ${\bf k}$ is the transverse momentum of the photon. In agreement with the experiment $k_{z}$ is $\pm k_{\gamma}$, since the frequency of the pump laser is such that in the longitudinal direction the absorption of photons with mode number $q =7$  dominates over other absorption processes \cite{BECphoton}. For the molecules we take an ideal gas in a box with volume $V$. Since this gas is at equilibrium and at room temperature, we describe the translational motion of the molecules by a classical Maxwell-Boltzmann distribution. Furthermore, we model these molecules as a two-level system with energy difference $\Delta > 0$ between the excited and ground state. This is a simplification since these molecules have a rovibritional structure. Therefore, the dye molecules have more than two levels as the rovibritional structure divides the ground and excited level into several sublevels. However, in this section we will show that we can model this multi-level system by introducing an effective mass for the molecules in our two-level model. 
\newline
\indent At time $t_{0}$ the photons are coupled to the molecules with a momentum-independent coupling constant $g$. To study the dynamics of the coupled system at times larger than $t_{0}$ we consider the action 
\begin{align}\label{eq:action}
S[&a_{{\bf k}},a^{*}_{{\bf k}},b_{{\bf k}},b^{*}_{{\bf k}}] = \\ \nonumber
&\, \sum_{{\bf k}} \int_{G^{\infty}} dt \, a_{{\bf k}}^{*}(t) \left\{i \hbar \frac{\partial}{\partial t} - \epsilon_{\gamma}({\bf k}) + \mu_{\gamma} \right\} a_{{\bf k}}(t) \\ \nonumber
+&\, \sum_{{\bf p},\rho} \int_{G^{\infty}} dt \, b_{{\bf p},\rho}^{*}(t) \left\{i \hbar\frac{\partial}{\partial t} - \epsilon({\bf p}) + \mu_{\rho} - K_{\rho} \right\} b_{{\bf p},\rho}(t) \\ \nonumber
-&\, \frac{i}{\sqrt{2V}} \sum_{{\bf k},{\bf p}} \int_{G^{\infty}} dt \, g a_{{\bf k}}(t) b_{{\bf p},\downarrow}(t) b^{*}_{{\bf p} + {\bf k}_{+},\uparrow}(t) + \mathrm{h.c.} \\ \nonumber
+&\, \frac{i}{\sqrt{2V}} \sum_{{\bf k},{\bf p}} \int_{G^{\infty}} dt \, g a_{{\bf k}}(t) b_{{\bf p},\downarrow}(t) b^{*}_{{\bf p} + {\bf k}_{-},\uparrow}(t) + \mathrm{h.c.}.
\end{align}
\noindent Here time is integrated along the Schwinger-Keldysh contour $G^{\infty}$, which is depicted in Fig. \ref{fig:SK}. The photons are described by the fields $a_{{\bf k}}(t)$ and $a^{*}_{{\bf k}}(t)$. Furthermore, $\epsilon_{\gamma}({\bf k})$ is given by Eq. \eqref{eq:enphot} and $\mu_{\gamma}$ is the chemical potential of the photons. For now we neglect the harmonic potential for the photons, since this term is not important for the coupling between molecules and photons. The fields $b_{{\bf p},\rho}(t)$ and $b^{*}_{{\bf p},\rho}(t)$ describe the dye molecules, with $\rho$ equal to $\downarrow$ or $\uparrow$, corresponding to the ground or excited state, respectively. Also, $\epsilon({\bf p}) = \hbar^{2} {\bf p}^{2}/2 m_{\mathrm{d}}$ with $m_{\mathrm{d}}$ the mass of the Rhodamine 6G molecule. Moreover, $K_{\rho}$ accounts for the energy difference between the molecular states and we take $K_{\downarrow} = 0$ and $K_{\uparrow} = \Delta$. The last two terms describe the process of respectively the absorption and emission of a photon. Here $g$ is the coupling strength between the photons and molecules, ${\bf k}_{+} = (k_{{\mathrm{x}}},k_{{\mathrm{y}}},k_{\gamma})$ and ${\bf k}_{-} = (k_{{\mathrm{x}}},k_{{\mathrm{y}}},-k_{\gamma})$. Note that the structure of the interaction terms is a consequence of the expansion of the photon field in terms of a standing wave, instead of a plane wave, in the z-direction. Furthermore, the summation over ${\bf k}$ is two-dimensional, wheras the summations over ${\bf p}$ are three-dimensional. The latter convention will be used throughout the paper.
\newline
\indent In this system one of the two molecular chemical potentials determines the density of molecules. In the experiment of Klaers \emph{et al.} one used Rhodamine 6G dye solved in methanol of concentration $1.5 \cdot 10^{-3} \, \mathrm{mol} \cdot \mathrm{L^{-1}}$. Therefore, we use a typical value of $n_{\mathrm{m}} = 9 \cdot 10^{23} \, \mathrm{m}^{-3}$ for the density of molecules. Furthermore the value of $\Delta\mu = \mu_{\uparrow} - \mu_{\downarrow}$ determines the polarization of the molecules. This polarization is defined as
\begin{align}\label{eq:pola}
P(\Delta\mu) := \frac{N_{\uparrow} - N_{\downarrow}}{N_{\uparrow} + N_{\downarrow}}= \frac{e^{\beta(\Delta\mu - \Delta) } - 1}{e^{\beta(\Delta\mu - \Delta) } + 1},
\end{align}
where $\beta$ is the inverse of the thermal energy $k_{\mathrm{B}}T$, and $N_{\uparrow}$ and $N_{\downarrow}$ are respectively the total number of excited-state and ground-state molecules. For small $\Delta\mu$ all molecules are in the ground state. By increasing the value of $\Delta\mu$, the number of molecules in the excited state increases. Since the total number of molecules is constant, the number of ground-state molecules thereby decreases. Thus for increasing $\Delta\mu$ the polarization increases. Moreover, the polarization is exactly zero for $\Delta\mu = \Delta$. A plot of the polarization as a function of $\Delta\mu$ is given in Fig. \ref{fig:polarization}. The parameter $\Delta\mu$ is also important for making a connection with the experiment, since the number of excited molecules and thereby the polarization is determined by the pumping power of the external laser. 
\newline
\indent
In our non-equilibrium theory the chemical potential of the photons $\mu_{\gamma}$ becomes only well defined after the photon gas equilbrates by coupling to the dye molecules. Since both the sum of the number of ground-state molecules and excited-state molecules, and the sum of the number of excited-state molecules and photons is constant, we have in equilibrium
\begin{eqnarray}\label{eq:chempot}
\Delta\mu &=& \mu_{\gamma}.
\end{eqnarray}
To derive an effective action for the photons, we first integrate out the molecules. 
\noindent Next we use perturbation theory up to second order in $g$ to obtain
\begin{align}\label{eq:keldac}
&S^{\mathrm{eff}}[a_{{\bf k}},a^{*}_{{\bf k}}] = \sum_{{\bf k}} \int_{G^{\infty}} dt^{\prime} \int_{G^{\infty}} dt \, a_{{\bf k}}^{*}(t) \\ \nonumber
&\, \times \left(\left\{i \hbar \frac{\partial}{\partial t} - \epsilon_{\gamma}({\bf k}) + \mu_{\gamma} \right\}\delta(t,t^{\prime}) - \hbar \Sigma({\bf k}, t,t^{\prime}) \right)a_{{\bf k}}(t^{\prime}), 
\end{align}
where the photon self-energy due to coupling with the dye is given by
\begin{align}\label{eq:self}
\Sigma({\bf k}, t,t^{\prime})  &= \frac{-i |g|^{2}}{2 \hbar^{2} V} \sum_{{\bf p}}G_{\downarrow}({\bf p},t^{\prime},t) \\ \nonumber
&\times \Big{\{} G_{\uparrow}({\bf k}_{+} + {\bf p},t,t^{\prime}) + G_{\uparrow}({\bf k}_{-} + {\bf p},t,t^{\prime}) \Big{\}}.
\end{align}
It turns out that both terms in the right-hand side are equal, and therefore we can write
\begin{align}
\Sigma({\bf k}, t,t^{\prime})= \frac{-i |g|^{2}}{\hbar^{2} V} \sum_{{\bf p}}G_{\downarrow}({\bf p},t^{\prime},t)G_{\uparrow}({\bf k}_{+} + {\bf p},t,t^{\prime}).
\end{align}
Here the Keldysh Green's function for the dye molecules is given by
\begin{align}\label{eq:gmb}
G_{\rho}({\bf p},t,t^{\prime}) &= i e^{-i (\epsilon({\bf p}) - \mu_{\rho} + K_{\rho}) (t -t^{\prime})/ \hbar} \\ \nonumber
&\times \{\Theta(t,t^{\prime}) (N_{\rho}({\bf p}) -1) + \Theta(t^{\prime},t) N_{\rho}({\bf p})\},
\end{align}
where $\Theta(t,t^{\prime})$ and $\Theta(t^{\prime},t)$ are the corresponding Heaviside functions on the Schwinger-Keldysh contour. Furthermore, the occupation numbers for the dye molecules are
\begin{eqnarray}
N_{\rho}({\bf p}) = e^{-\beta (\epsilon({\bf p}) - \mu_{\rho} + K_{\rho})},
\end{eqnarray}
with $\rho \in \{\uparrow,\downarrow\}$. Since this action is defined on the Schwinger-Keldysh contour, we can only use this action to calculate quantities on the Schwinger-Keldysh contour. However, the relevant physical quantities should be calculated on the real-time axis. Therefore, we need to transform this action into an action that is defined on this real-time axis. As is shown in Ref. \cite{Stoofkeldysh}, this boils down to determing the retarded, advanced and Keldysh self-energy. Roughly speaking, the advanced and retarded self energies determine the dynamics of the single-particle wavefunction in the gas, i.e., the coherent dynamics, and the Keldysh component accounts for the dynamics of their occupation numbers, i.e., the incoherent dynamics. 
\newline
\indent In the continuum limit the retarded self-energy becomes
\begin{align}\label{eq:selfplus}
\hbar \Sigma^{(+)} &({\bf k}, t - t^{\prime}) = \frac{i}{\hbar} \Theta(t - t^{\prime}) \int \frac{d {\bf p}}{(2 \pi)^{3}} \, |g|^{2} \\ \nonumber
&\times e^{i (\epsilon({\bf k}_{+},{\bf p}) + \Delta\mu) (t -t^{\prime})/ \hbar} \Big{\{} N_{\uparrow}({\bf k}_{+} + {\bf p}) - N_{\downarrow}({\bf p}) \Big{\}}.
\end{align}
\noindent Here we used Eqs. \eqref{eq:self} and \eqref{eq:gmb} and we defined $\epsilon({\bf k}_{+},{\bf p}) = \epsilon({\bf p}) - \epsilon({\bf k}_{+} + {\bf p}) - \Delta$. In Fourier space this self-energy reads
\begin{align}\label{eq:selfplusomega}
\hbar \Sigma^{(+)} ({\bf k}, \omega) &:= S({\bf k}, \omega) - i R({\bf k}, \omega)  \\ \nonumber
&:= \int d(t - t^{\prime}) \, \hbar \Sigma^{(+)} ({\bf k}, t - t^{\prime}) e^{i \omega (t - t^{\prime})}.
\end{align}
Since the molecules behave as a Maxwell-Boltzmann gas at room temperature, we can find an analytical expression for $R({\bf k}, \omega)$. We obtain
\begin{align}\label{eq:imself}
R({\bf k}, \omega) &= A({\bf k}, \omega) \frac{|g|^{2} m^{2}_{\mathrm{d}}}{2 |{\bf k}_{+}| \pi \beta \hbar^{4}} \sinh\Bigg{\{}\frac{\beta \hbar \omega}{2}\Bigg{\}},
\end{align}
with
\begin{align}\label{eq:imself2}
A({\bf k}, \omega) &=  \exp \Big{\{} \beta \Big{(}\mu_{\downarrow} + \mu_{\uparrow} - \Delta \Big{)}/2 \Big{\}} \\ \nonumber
&\times \exp \Bigg{\{} - \frac{\beta}{4} \Bigg{[}\epsilon({\bf k}_{+})  + \frac{(\Delta - \Delta\mu - \hbar \omega)^{2}}{\epsilon({\bf k}_{+})}\Bigg{]} \Bigg{\}}.
\end{align}
Furthermore, in Fourier space the Keldysh self-energy is given by
\begin{align}\label{eq:sigmak}
&\hbar \Sigma^{K}({\bf k}, \omega) = i \int \frac{d {\bf p}}{(2 \pi)^{2}} \, \delta(\hbar \omega + \epsilon({\bf k}_{+},{\bf p}) + \Delta\mu) \\ \nonumber
&\, \times |g|^{2} \{2 N_{\downarrow}({\bf p}) N_{\uparrow}({\bf k}_{+} + {\bf p}) - N_{\downarrow}({\bf p}) - N_{\uparrow}({\bf k}_{+} + {\bf p})\}.
\end{align}
Since the dye is in quasi-equilibrium, this Keldysh self-energy can be related to the imaginary part of the retarded self-energy. We find
\begin{eqnarray}\label{eq:flucdis}
\hbar \Sigma^{K}({\bf k}, \omega)  = -2i (1 + 2 N(\omega))R({\bf k}, \omega),
\end{eqnarray}
where 
\begin{eqnarray}
N(\omega) = \frac{1}{e^{\beta \hbar \omega} - 1}.
\end{eqnarray}
This result is known as the fluctuation-dissipation theorem. As we show in the next section, this result guarantees that the photon gas relaxes towards thermal equilibrium in the limit of $t \rightarrow \infty$.
\newline
\indent To make further progress, we have to determine typical numerical values for $\Delta$ and $g$ appropriate for the experiment of Klaers \emph{et al.}. These values can be obtained by looking at the physical meaning of the self-energy. Consider a system of molecules that can be either in a ground state or excited state. If we apply a laser to this system, we can measure for instance the total number of molecules in the excited state. This number depends on the rate of photon absorption and emission, and therefore on the lifetime of the photons. Since the imaginary part of the retarded self-energy is related to the lifetime of the photon, we can determine the emission and absorption spectrum of the molecules with the help of our expression for $R({\bf k}, \omega)$. 
\newline
\indent In order to obtain the absorption and emission spectrum seperately, we take a closer look at the retarded self-energy given by Eq. \eqref{eq:selfplus}. In this expression the factor with the Maxwell-Boltzmann distribution can be rewritten as $N_{\downarrow}({\bf p}) (N_{\uparrow}({\bf k}_{+} + {\bf p}) \pm 1) - N_{\uparrow}({\bf k}_{+} + {\bf p})(N_{\downarrow}({\bf p}) \pm 1)$. The first term can be understood as the absorption of a photon, since this statistical factor accounts for the process where a ground-state molecule scatters into an excited state. The factor $N_{\downarrow}({\bf p})$ simply is the number of molecules that can undergo the collision and $(N_{\uparrow}({\bf k}_{+} + {\bf p}) \pm 1)$ denotes the Bose enhancement factor or Pauli blocking factor depending on the quantum statistics of the dye molecules. By using a similar reasoning the second term can be understood as the emission of a photon. Hence, the part of the self-energy proportional to $N_{\downarrow}({\bf p})$ is related to the absorption spectrum, and the part proportional to $N_{\uparrow}({\bf k}_{+} + {\bf p})$ is related to the emission spectrum. 
\newline
\indent The absorption and emission spectra are usually obtained in experiments where the number of photons is not conserved. So, in these systems the photons have no chemical potential. To make a comparison, we therefore have to set $\mu_{\downarrow} = \mu_{\uparrow}$. Furthermore, contrary to the experiment of Klaers \emph{et al.}, there is no restriction on the momentum of the photons. This implies that the photon field should be expanded into plane waves instead of standing waves. Therefore, the fourth term in the right-hand side of the action in Eq. \eqref{eq:action} is absent and the prefactor of the third term is changed into $1/\sqrt{V}$. However, this modification leaves the expressions for the self-energies unchanged.
\newline
\indent In order to get more insight into the role of the parameters of our model in the absorption and emission spectrum, we first consider the experiment of Klaers \emph{et al.} and we keep ${\bf k}$ fixed. Then, the spectra have a maximum at 
\begin{align}\label{eq:max1}
\hbar \omega_{\pm} = \Delta \pm \frac{\hbar^{2} {\bf k}_{+}^{2}}{2 m_{\mathrm{d}}},
\end{align}
where we used that $\Delta\mu = 0$. Here the plus sign is the maximum of the absorption spectrum and the minus sign corresponds to the position of the maximum of the emission spectrum. So we obtain a difference in frequency between the maximum of the absorption and emission spectrum. This difference is also obtained experimentally and known as the Stokes shift. From this expression we find that the value of $m_{\mathrm{d}}$ determines the value of the Stokes shift. Furthermore we can see from Eq. \eqref{eq:max1}, that we can change the position of the peaks by varying $\Delta$. 
\begin{figure}[t]
 \includegraphics{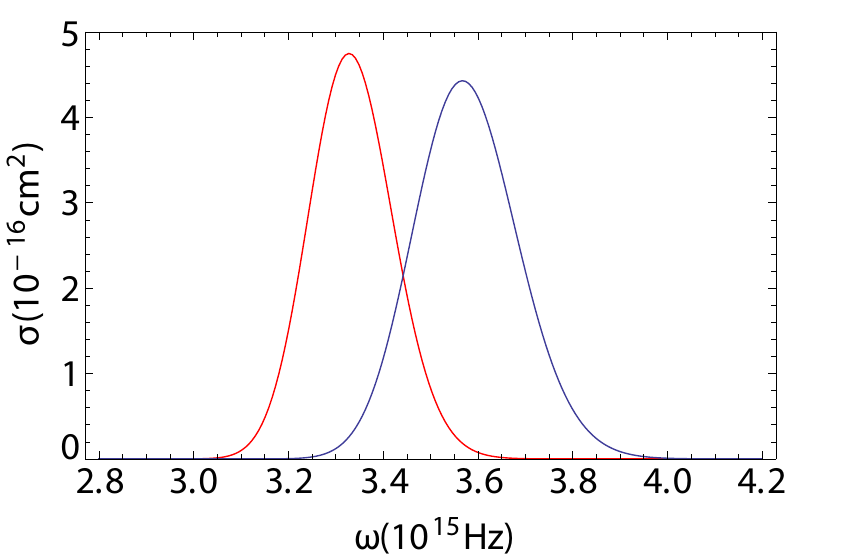}
\vspace{-0.5 cm}
 \caption{The absorption and emission cross section of the photons obtained from the imaginary part of the retarded self-energy for $m_{\mathrm{d}} = 9.3 \cdot 10^{-35}$ kg, $\Delta = 3.63 \cdot 10^{-19} \, \mathrm{J}$ and $g = 6.08 \cdot 10^{-26} \, \mathrm{J} \cdot \mathrm{m^{3/2}}$. The red curve corresponds to the emission and the blue curve denotes the absorption of photons. The absorption cross section is given by Eq. \eqref{eq:abs}, and the plotted emission cross section is obtained from the same equation by replacing $R_{\mathrm{abs}}(\omega)$ by $R_{\mathrm{emis}}(\omega)$.} 
 \label{fig:AbsEmis}
\end{figure}
\newline
\indent Now we turn to more conventional experiments, where the absorption of laser light by the medium is measured as a function of frequency. To obtain the absorption and emission spectra only as a function of frequency, we have to consider the self-energy on shell and thus replace ${\bf k_{+}}$ by $\omega/c$. For the physical mass of the Rhodamine 6G molecule, we obtain peaks that are too narrow and a Stoke shift that is too small. This is because we neglected the rovibrational structure of the molecules. Due to this rovibrational structure there are many possible transitions, since the excited and ground level are split into several sublevels. Therefore there is a whole range of photon energies which can be absorpted or emitted by the molecule. This causes a considerable broadening of the spectra. As mentioned before, we model this rovibritional structure by choosing an effective mass for the molecules. We can also see explicitly from Eqs. \eqref{eq:imself} and \eqref{eq:imself2}, that decreasing the value of $m_{\mathrm{d}}$ will indeed broaden the peaks. For $\Delta = 3.63 \cdot 10^{-19}$ J and $ m_{\mathrm{d}} = 9.3 \cdot 10^{-35}$ kg we recover in good approximation the normalized absorption and emission spectrum given in Ref. \cite{BECphoton}. 
\newline
\indent Up to now, we have considered the relative absorption and emission spectrum. To obtain the correct height of these spectra, we have to find an appropiate value for $g$. By using Ref. \cite{Spectrum}, we can actually compare our results to the experimentally obtained absorption and emission spectra. However, to calculate the emission spectrum for this particular
experiment within our formalism, we have to take into account that the emission of a photon can be in an arbitrary direction. Thus to obtain the emission spectrum we have to perform an integral which averages over all possible emission directions. However, the absorption spectrum can be obtained without performing additional integrals and therefore we focus on this spectrum to obtain a numerical value for $g$. Then, as a consequence of our formalism also the for our purposes correct emission spectrum is incorporated. 
\newline
\indent Before we can fit $g$, we have to relate our calculated decay rates to the measured absorption cross sections. We have
\begin{eqnarray}
\frac{dN}{dx} = - n_{\downarrow} \sigma_{\mathrm{abs}}(\omega),
\end{eqnarray}
\noindent
where the left hand-side is the number of absorbed photons $dN$ in a distance $dx$ along the path of a beam. Furthermore, $n_{\downarrow}$ is the density of ground-state molecules and $\sigma_{\mathrm{abs}}$ is the absorption cross section. By using Fermi's golden rule, we obtain that $dN/dt$ is equal to $-2 R_{\mathrm{abs}}(\omega)/\hbar$. Here $R_{\mathrm{abs}}(\omega)$ denotes the absorption term in the imaginary part of the self energy. Hence
\begin{eqnarray}\label{eq:abs}
\sigma_{\mathrm{abs}}(\omega) = \frac{2 R_{\mathrm{abs}}(\omega)}{c \hbar n_{\downarrow}}.
\end{eqnarray}
Since the molecules behave as a classical Maxwell-Boltzmann gas,
\begin{eqnarray}\label{eq:densities}
n_{\downarrow} = \left(\frac{m_{\mathrm{d,real}}}{2 \pi \hbar^{2} \beta} \right)^{3/2} e^{\beta \mu_{\downarrow}}.
\end{eqnarray}
Note that contrary to the mass of the dye molecules used in the self-energies, we here use the real mass of the dye molecules to obtain the correct densities. Thus $m_{\mathrm{d}}$ is the effective mass to model the rovibrational structure of the molecules and $m_{\mathrm{d,real}} \simeq 7.95 \cdot 10^{-25}$ kg is the physical mass of a Rhodamine 6G molecule. By using our expression for $R({\bf k}, \omega)$ we observe that the absorption cross section is independent of $\mu_{\downarrow}$. Therefore we do not need to specify the number of molecules to obtain a numerical estimate for $g$. Furthermore, we can relate the absorption cross section given by Eq. \eqref{eq:abs} to the molecular extinction coeffient obtained in Ref. \cite{Spectrum}. According to Ref. \cite{Spectrum2},
\begin{align}
\sigma = (3.82 \cdot 10^{-21} \,\mathrm{cm}^{3} \cdot \mathrm{mol} \cdot \mathrm{L}^{-1}) \cdot \epsilon,
\end{align}
where $\epsilon = 1.16 \cdot 10^{5} \, \mathrm{L} \cdot \mathrm{mol}^{-1} \cdot \mathrm{cm}^{-1}$ is the molar extinction coefficient. This results into $g \simeq 6.08 \cdot 10^{-26} \, \mathrm{J} \cdot \mathrm{m^{3/2}}$. A plot of the absorption and emission cross section for the obtained numerical values for $\Delta$, $m_{\mathrm{d}}$ and $g$ is given in Fig. \ref{fig:AbsEmis}. The shown emission cross section is obtained from Eq. \eqref{eq:abs} by replacing $R_{\mathrm{abs}}(\omega)$ by $R_{\mathrm{emis}}(\omega)$. As mentioned before, this is not the physical emission cross section since that can only be obtained by integrating over all directions of emission.

\section{Non-equilibrium physics}
\label{sec:nonequi}
\begin{figure}[h!]
 \centerline{\includegraphics{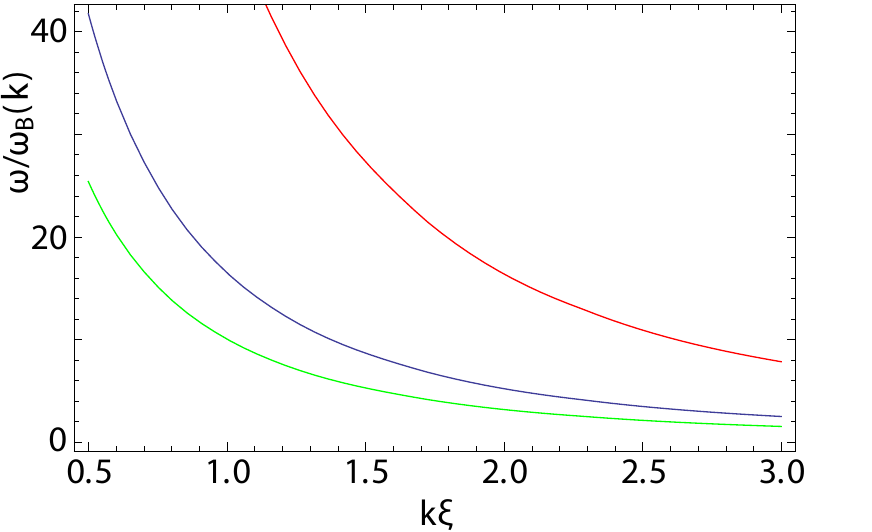}}
 \caption{A plot of the validity of the linear approximation of the retarded self-energy for $n_{{0}} = 10^{12}  \, \mathrm{m}^{-2}$ and certain values of $\omega$ and $k$. The blue, red and green curve are for respectively $\Delta\mu$ equal to $3.4 \cdot 10^{-19} \, \mathrm{J}$, $3.7 \cdot 10^{-19} \, \mathrm{J}$ and $4.0 \cdot 10^{-19} \, \mathrm{J}$. Below the curves is the region where the linear approximation is within $1 \%$ of the actual value of the self-energy.} 
 \label{fig:selfapprox}
\end{figure}

\begin{figure}[h!]
\centerline{\includegraphics{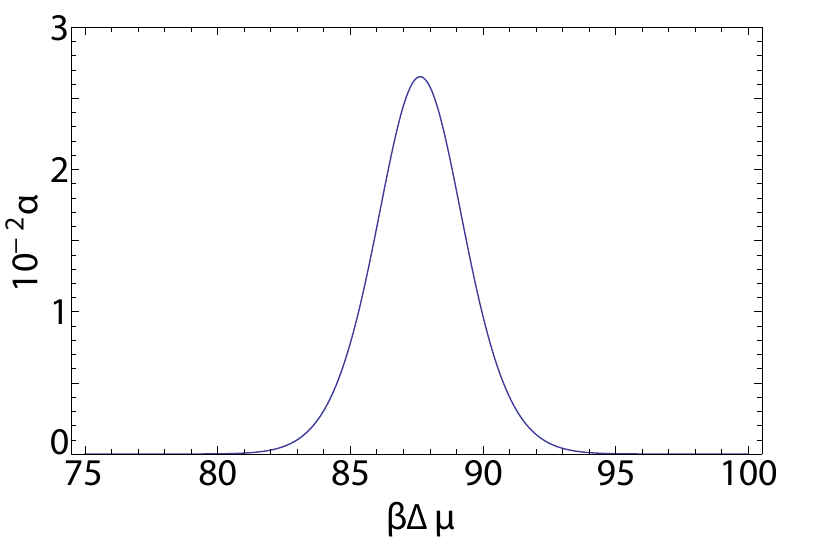}}
\vspace{-0.1 cm}
\caption{A plot of the dimensionless damping parameter $\alpha$ as a function of $\beta\Delta\mu$. For this plot we used $n_{\mathrm{m}} = 9 \cdot 10^{23}  \mathrm{m}^{-3}$. The parameter $\alpha$ has a maximum value of about $2.65 \cdot 10^{-2}$ at $\Delta\mu$ equal to $\Delta$.} 
 \label{fig:alpha}
\end{figure}

We introduce a complex field $\phi({\bf x},t)$ for the photons such that
\begin{align}\label{eq:dens}
\langle |\phi({\bf k}, t)|^{2} \rangle = N({\bf k}, t) + \frac{1}{2},
\end{align}
where $N({\bf k}, t)$ corresponds to the average occupation number of the single-particle state with momentum ${\bf k}$ at time $t$. As is shown in Ref. \cite{Stoofkeldysh}, $\phi({\bf x},t)$ obeys a Langevin field equation for describing the dynamics of the photon gas. This equation ultimately reads
\begin{align}
i \hbar &\frac{\partial}{\partial t} \phi({\bf x},t) = \Big{(} H({\bf x}) + T |\phi({\bf x},t)|^{2}  \Big{)}  \phi({\bf x},t) \\ \nonumber
&+ \int d{\bf x}^{\prime} dt^{\prime} \hbar\Sigma^{(+)}({\bf x} - {\bf x}^{\prime},t - t^{\prime})  \phi({\bf x}^{\prime},t^{\prime}) +  \eta({\bf x},t),
\end{align}
where the Hamiltonian
\begin{align}
H({\bf x}) = -\frac{\hbar^{2} \nabla^{2}}{2 m_{\mathrm{ph}}} - \mu_{\gamma} + \hbar c k_{\gamma} + \frac{1}{2}m_{\mathrm{ph}} \Omega^{2} |{\bf x}|^{2}.
\end{align}
Here ${\bf x} = (x,y)$ and the field $\phi^{*}({\bf x},t)$ satisfies the complex conjugate equation. Furthermore, to obtain this equation we expanded Eq. \eqref{eq:enphot} for small transverse momenta, using that $k_{\gamma}({\bf x})$ is position dependent due to the curvature of the cavity mirrors. In this equation $m_{\mathrm{ph}} \simeq 6.7 \cdot 10^{-36} \, \mathrm{kg}$ is the effective mass of the photons and $\Omega \simeq 2.6 \cdot 10^{11} \, \mathrm{Hz}$ is the trapping frequency of the harmonic potential. We also introduced a self-interaction term with strength $T \simeq 1.2 \cdot 10^{-36} \, \mathrm{J \cdot m^{2}}$. According to Ref. \cite{BECphoton}, this self-interaction of the photons arises from Kerr nonlinearity or thermal lensing in the dye. Finally, the Gaussian noise $\eta({\bf x},t)$ satisfies  
\begin{align}\label{eq:fluc33}
\langle \eta({\bf x},t) \eta^{*}({\bf x}^{\prime},t^{\prime}) \rangle = \frac{i \hbar}{2} \hbar\Sigma^{K}({\bf x} - {\bf x}^{\prime}, t - t^{\prime}).
\end{align}
Here, the brackets denote averaging over different realizations of the noise. In general it is difficult to determine correlation functions from these equations, especially because of the non-locality of the retarded self-energy. However, since we are interested in Bose-Einstein condensation of photons we focus on the low-energy behaviour of this self-energy. In the low-energy regime we are interested in $k \xi$ of the order of unity and $\omega$ around $\omega_{\mathrm{B}}({\bf k})$, where
\begin{align}
\hbar \omega_{\mathrm{B}}({\bf k}) = \sqrt{\left(\frac{\hbar^{2} {\bf k}^{2}}{2 m_{\mathrm{ph}}}\right)^{2} + 2 n_{0} T \left(\frac{\hbar^{2} {\bf k}^{2}}{2 m_{\mathrm{ph}}}\right)},
\end{align}
is the Bogoliubov dispersion and
\begin{align}
\xi = \frac{\hbar}{2 \sqrt{m_{\mathrm{ph}} n_{0} T}},
\end{align}
is the coherence length, with $n_{0}$ the density of condensed photons. Note that $k$ is the norm of the two-dimensional momentum vector ${\bf k} = (k_{\mathrm{x}},k_{\mathrm{y}})$. 
Since for the experiment of Ref. \cite{BECphoton} the critical number of photons $N_{\mathrm{c}} \simeq 77000$ and the diameter of the condensate is measured as a function of the condensate fraction, we can make an estimate for $n_{{0}}$. We obtain condensate densities in the range of at least $10^{12} - 10^{13}  \, \mathrm{m}^{-2}$.
\newline
\indent We make a low-energy approximation to the imaginary part of the retarded self energy. As we see in Fig. \ref{fig:selfapprox}, this is a good approximation in the low-energy regime. Furthermore, the real part of the retarded self-energy is small and as a zeroth-order approximation we neglect this contribution. Thus we approximate
\begin{align}\label{eq:selfap}
\hbar\Sigma^{(+)}({\bf k}, \omega) = - i \alpha \hbar \omega,
\end{align}
and we can write for the Langevin field equation that
\begin{align}\label{eq:lagalpha}
i \hbar (1 + i\alpha) \frac{\partial}{\partial t} \phi({\bf x},t) &= (H({\bf x}) + T |\phi({\bf x},t)|^{2})\phi({\bf x},t) \\ \nonumber
&+ \eta({\bf x},t).
\end{align}
This is the equation which determines the complete dynamics of the photon gas. The finite lifetime effects are captured by the single dimensionless parameter $\alpha$, that depends on the difference between the chemical potentials of the excited-state and ground-state molecules. Furthermore, the noise $\eta({\bf x},t)$ is related to the Keldysh self-energy via Eq. \eqref{eq:fluc33} and in this approximation
\begin{align}
\hbar\Sigma^{K}&({\bf x}^{\prime} - {\bf x}, t^{\prime} - t) = - i \delta({\bf x} - {\bf x}^{\prime}) \\ \nonumber
&\times 2 \alpha\hbar  \int \frac{d\omega}{2 \pi} \left(1 + 2 N(\omega) \right) \omega  e^{-i \omega (t^{\prime} - t)}.
\end{align}
The explicit dependence of $\alpha$ on $\Delta\mu$ is given by 
\begin{align}\label{eq:alphachem}
\alpha =  \alpha_{\mathrm{max}} n_{\mathrm{m}} \frac{e^{-C (\Delta\mu - \Delta)^{2}}}{\cosh\left\{{\frac{1}{2}\beta (\Delta\mu - \Delta)}\right\}},
\end{align}
where $n_{\mathrm{m}}$ is the density of dye molecules,
\begin{align}
C = \frac{\beta m_{\mathrm{d}}}{2 \hbar^{2} |k_{\gamma}|^{2}},
\end{align} 
and
\begin{align}
\alpha_{\mathrm{max}} = \sqrt{\frac{\pi m_{{\mathrm{d,real}}}}{8 \beta \hbar^{2} k_{\gamma}^{2}}} \left(\frac{\beta |g| m_{\mathrm{d}}}{m_{{\mathrm{d,real}}}}\right)^{2} e^{-\beta \hbar^{2} k_{\gamma}^{2}/ 8 m_{\mathrm{d}}}.
\end{align}
The damping parameter $\alpha$ is inversely proportional to the photon lifetime, and accounts for the decay of photons due to the interaction with the dye molecules. The emission and absorption of photons are equally important for the photon equilibration. Therefore, $\alpha$ has a maximum when there is an equal amount of excited-state and ground-state molecules, i.e., for $\Delta\mu = \Delta$ where $P = 0$. This also explains the symmetric form of $\alpha$ around $\Delta\mu = \Delta$. Namely, $\alpha$ should be symmetric under changing the sign of the polarization as this only interchanges the excited-state and ground-state molecule densities. A plot of $\alpha$ as a function of $\Delta\mu$ is shown in Fig. \ref{fig:alpha}. 
\newline
\indent The Langevin field equation given by Eq. \eqref{eq:lagalpha} incorporates the complete dynamics of the photons. 
\noindent However, we still need to check that for large times the photon distribution function relaxes to the correct equilibrium. For this it suffices to consider the homogeneous case and to neglect the self-interaction of the photons. For purpose of generality, we do not make a low-energy approximation to the self-energy and we Fourier transform the Langevin field equation into
\begin{align}\label{eq:equil2}
i \hbar \frac{\partial}{\partial t} \phi({\bf k}, t) &= (\epsilon_{\gamma}({\bf k}) - \mu_{\gamma}) \phi({\bf k}, t)  + \eta({\bf k}, t) \\ \nonumber
&+ \int_{t_{0}}^{\infty} dt^{\prime} \, \hbar\Sigma^{(+)}({\bf k}, t - t^{\prime}) \phi({\bf k}, t^{\prime}).
\end{align}
\indent As mentioned in the previous section, the fluctuation-dissipation theorem given by Eq. \eqref{eq:flucdis} should insure that the gas relaxes towards thermal equilibrium. To check that this formalism contains this correct equilibrium, we assume that $\langle \phi({\bf k}, t)  \phi^{*}({\bf k}, t^{\prime}) \rangle$ only depends on the difference $t - t^{\prime}$ and write,
\begin{align}\label{eq:bose}
\langle \phi({\bf k}, t)  \phi^{*}({\bf k}, t^{\prime}) \rangle = \int \frac{d\omega}{2 \pi} G(\omega) e^{-i \omega (t - t^{\prime})}.
\end{align}
Then
\begin{align}\label{eq:equil}
i \hbar \frac{d}{d (t + t^{\prime})} \langle \phi({\bf k}, t)  \phi^{*}({\bf k}, t^{\prime}) \rangle = 0,
\end{align}
and for $t^{\prime} = t$ we obtain the equilibrium value for $\langle \phi({\bf k}, t)  \phi^{*}({\bf k}, t) \rangle$. Since we are interested in equilibrium, we consider times much larger than $t_{0}$. Therefore we are allowed to take the limit of $t_{0} \rightarrow -\infty$. Now Eq. \eqref{eq:equil} can be rewritten as
\begin{align}\label{eq:corr2}
&\langle \eta({\bf k}, t) \phi^{*}({\bf k}, t^{\prime}) \rangle -  \langle \phi({\bf k}, t) \eta^{*}({\bf k}, t^{\prime}) \rangle = \\ \nonumber
&\int_{-\infty}^{\infty} dt^{\prime\prime} \, \langle  \phi({\bf k}, t) \phi^{*}({\bf k}, t^{\prime\prime}) \rangle  \hbar\Sigma^{(-)}({\bf k}, t^{\prime\prime} - t^{\prime}) - \\ \nonumber
&\int_{-\infty}^{\infty} dt^{\prime\prime} \, \hbar\Sigma^{(+)}({\bf k}, t - t^{\prime\prime}) \langle \phi({\bf k}, t^{\prime\prime}) \phi^{*}({\bf k}, t^{\prime}) \rangle.
\end{align}
Here we used that $\hbar \Sigma^{(-)} ({\bf k}, t^{\prime} - t) = (\hbar \Sigma^{(+)} ({\bf k}, t^{\prime} - t))^{*}$. Furthermore, since the field $\phi({\bf k}, t)$ and its complex conjugate depend on the noise, we have a non-zero value for $\langle \eta({\bf k}, t) \phi^{*}({\bf k}, t^{\prime}) \rangle$, which can be determined by formally integrating Eq. \eqref{eq:equil2} and using Eq. \eqref{eq:fluc33}. 
\noindent 
In Fourier space Eq. \eqref{eq:corr2} is given by
\begin{align}
-\frac{1}{2 i} \Sigma^{K}({\bf k}, \omega) G^{(+)} G^{(-)} =  G(\omega),
\end{align}
where the retarded (+) and advanced (-) photon Green's functions are determined by
\begin{align}\label{eq:gr1}
\hbar G^{(\pm),-1} = \hbar\omega^{\pm} -  \epsilon_{\gamma}({\bf k})  + \mu_{\gamma} - \hbar\Sigma^{(\pm)}({\bf k}, \omega).
\end{align}
To make further progress, we introduce the spectral function 
\begin{align}\label{eq:spectral}
\rho({\bf k},\omega) &= -\frac{1}{\pi\hbar} \mathrm{Im}\left[G^{(+)} \right] \\ \nonumber
&= \frac{1}{\pi} \frac{R({\bf k}, \omega)}{(\hbar \omega - \epsilon_{\gamma}({\bf k}) + \mu_{\gamma} - S({\bf k}, \omega))^{2} + (R({\bf k}, \omega))^{2}} \\ \nonumber
&= \frac{1}{\pi\hbar^{2}} R({\bf k}, \omega) G^{(+)} G^{(-)}.
\end{align}
This spectral function $\rho({\bf k},\omega)$ can be interpreted as a single-particle density of states. Therefore we can calculate densities in equilibrium by multiplying this spectral function with the Bose-distribution function $N(\omega)$ and then integrating over $\hbar \omega$. Hence,
\begin{align}
N({\bf k}) = \int d(\hbar\omega) \,  N(\omega) \rho({\bf k},\omega),
\end{align} 
where $N({\bf k})$ is the number of photons in a state with momentum ${\bf k}$. Thus, in equilibrium
\begin{align}
G(\omega) = 2\pi\hbar \left(\frac{1}{2} + N(\omega)\right) \rho({\bf k},\omega),
\end{align}
where we used the fluctuation-dissipation theorem in Eq. \eqref{eq:flucdis}. Hence,
\begin{align}
\langle \phi({\bf k}, t) \phi^{*}({\bf k}, t) \rangle = N({\bf k}) + \frac{1}{2}.
\end{align}
By comparing this result to Eq. \eqref{eq:dens}, we find that the average occupation numbers $N({\bf k}, t)$ relax to $N({\bf k})$. During this calculation we did not use an approximation for the imaginary part of the retarded self energy. However, we can do the same calculation for $R({\bf k}, \omega)$ given by $\alpha \hbar \omega$. This approximation will directly manifest itself in the fluctuation-dissipation theorem and ultimately in the spectral function. Therefore, also in this approximation the equilibrium occupation numbers are given by Bose-Einstein distribution functions.

\section{Equilibrium}
\label{sec:equil}
In the previous section we have shown that the complete dynamics of the photon gas can be obtained from a Langevin field equation for a complex field $\phi({\bf x},t)$. On top of this non-equilibrium physics, we demonstrated that the relaxation of the photons towards the correct equilibrium. 
\begin{figure*}[t]
  \begin{center}
\setlength{\abovecaptionskip}{17.5 pt plus 3pt minus 2pt}
    \mbox{
      \subfigure{\includegraphics{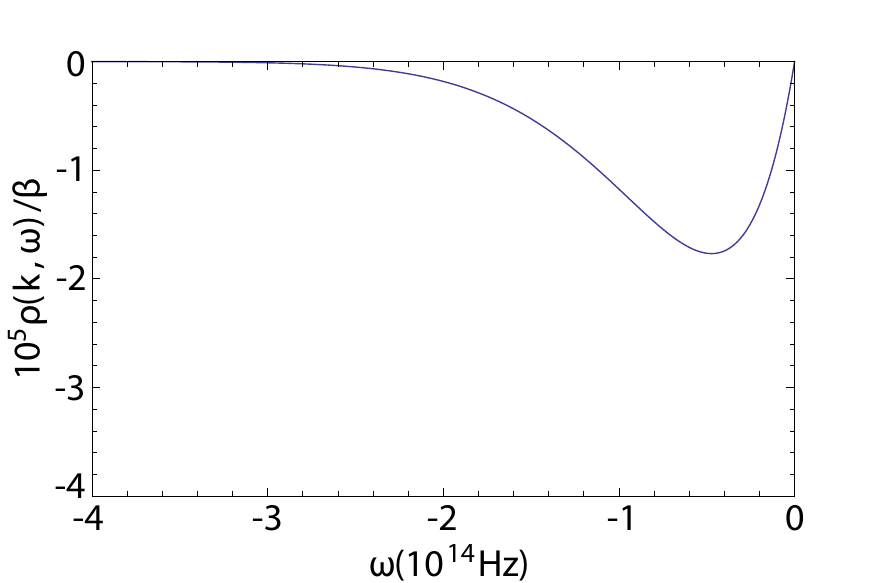}}  \,
      \subfigure{\includegraphics{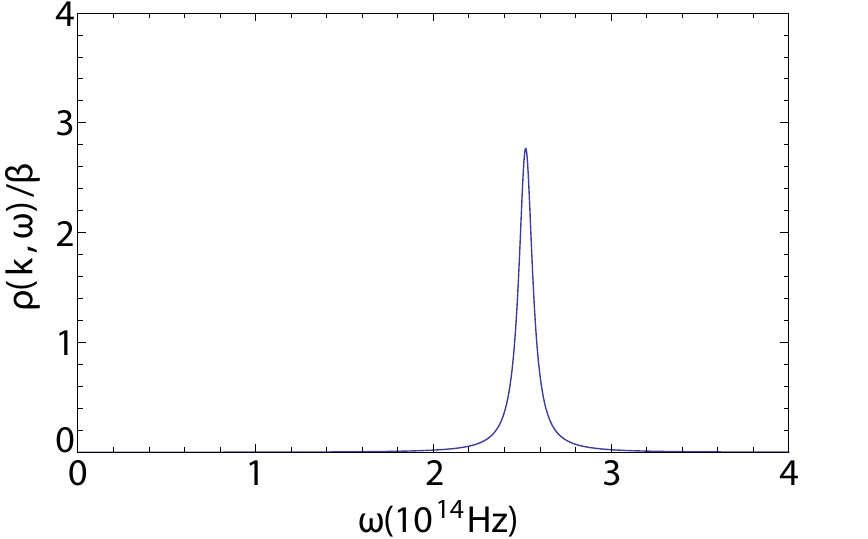}}
      }
\vspace{-0.6 cm}
    \caption{The spectral function as a function of the frequency $\omega$ for $k_{y} = k_{x} = 0$ for $n_{m} = 9 \cdot 10^{23} \mathrm{m}^{-3}$ and $\mu_{\gamma} = 3.5 \cdot 10^{-19} \mathrm{J}$. The negative (positive) contribution for negative (positive) frequencies is shown on the left (right)-hand side.}
    \label{fig:AMS}
  \end{center}
\end{figure*}
In this section we discuss equilibrium properties of the photon gas, and we therefore set $\Delta\mu = \mu_{\gamma}$ according to Eq. \eqref{eq:chempot}. We perform calculations both in the normal and in the Bose-Einstein condensed state. 

\subsection{Normal state}
We first consider the spectral function of the photons defined in Eq. \eqref{eq:spectral}. The spectral function should satisfy two conditions. First of all, because we are dealing with bosons, the spectral function should be positive (negative) for positive (negative) frequencies. From Eq. \eqref{eq:imself} it is clear that $R({\bf k}, \omega)$ has this property and therefore this condition is indeed satisfied by the spectral function. Secondly, the spectral function should satisfy the zeroth-frequency sum rule
\begin{align}\label{eq:sumrule}
\int d(\hbar \omega) \rho({\bf k},\omega) = 1.
\end{align}
By numerically integrating this spectral function, we check that we satisfy the sum rule for all chemical potential smaller than the lowest energy of the photons.
\newline
\indent As we can see from Fig. \ref{fig:AMS}, the spectral function consists of a Lorentzian-like peak for positive frequencies and a continuum for negative frequencies. The latter is roughly five orders of magnitude smaller than the positive contribution. Since the positive contribution is approximately a Lorentzian, we can determine the lifetime of the photons by looking at the width of these peaks \cite{UQF}. This lifetime is defined as the time for which a photon in a certain momentum state ${\bf k}$ goes into another state with momentum ${\bf k}^{\prime}$ due to absorption and re-emission by the molecules. 
\newline
\indent Numerically, we obtained for small momenta and $\beta \mu_{\gamma}$ up to roughly $87$, a lifetime on the order of $10^{-13}$ s. If we increase $\mu_{\gamma}$ even further, the lifetime of the photons increases rapidly. Because for larger values of $\mu_{\gamma}$ the peaks of the spectral function are at smaller frequencies, we can also show this fact analytically. Since we know that the lifetime of the photon is related to the imaginary part of this pole, we need to determine the poles of Eq. \eqref{eq:gr1}. 
\newline
\indent In the previous section we found that for small frequencies we can use an approximation for the imaginary part of the self-energy in which it is linear in frequency. Within this approximation, the Green's function given by Eq. \eqref{eq:gr1} has a pole at
\begin{align}
\hbar\omega^{\mathrm{pole}}({\bf k}) = \frac{1 - i \alpha}{1 + \alpha^{2}} \left(\epsilon_{\gamma}({\bf k}) - \mu_{\gamma}\right).
\end{align}
Since $\alpha^{2} \ll 1$, a typical lifetime of the photons in the normal state is given by
\begin{align}
\tau({\bf k}) = \frac{\hbar}{2 \alpha \left(\epsilon_{\gamma}({\bf k}) - \mu_{\gamma}\right)} \sim \frac{1}{\alpha \Omega}.
\end{align}
where $\alpha$ is given by Eq. \eqref{eq:alphachem}, $\Delta\mu$ is sufficiently large and $\Omega$ is the trap frequency of the photons. In the last step we used that the photons are trapped in a harmonic potential and therefore the typical energy of the photons is proportional to $\hbar\Omega$. From Fig. \ref{fig:alpha} we know that for the relevant values of $\mu_{\gamma}$, $\alpha$ is in the range of $10^{-3} - 10^{-2}$. Therefore, the lifetime of the photons is in the ns regime, which agrees with Ref. \cite{BECphoton}. We also note that the smallness of $\alpha$ implies that the collective-mode dynamics of the gas is underdamped as the ratio between the damping and frequency of the collective modes is precisely $\alpha$.

\subsection{Condensed state}
\begin{figure}[h!]
\includegraphics{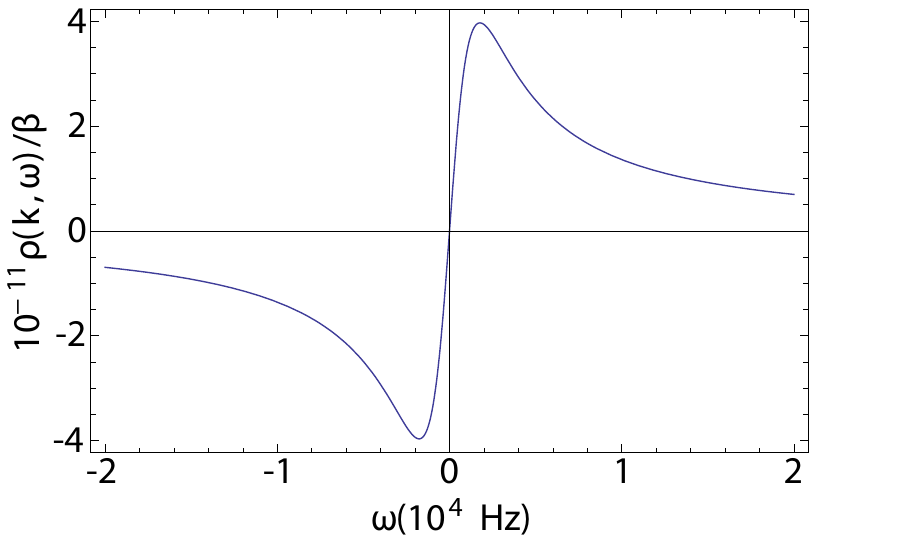}
 \caption{A plot of the spectral function as a function of $\omega$ for $n_{\mathrm{m}} = 9 \cdot 10^{23} \, \mathrm{m}^{-3}$ and $n_{0} = 10^{12} \,\mathrm{m}^{-2}$. In this plot $k \xi \simeq 1.9 \cdot 10^{-5}$.}
 \label{fig:Bogspectral2}
\end{figure}

In this subsection we consider the homogeneous two-dimensional photon gas below the critical temperature for Bose-Einstein condensation. To describe the condensate of photons we start from the following two-dimensional action
\begin{align}
S^{\mathrm{eff}}[a^{*},a]&= \sum_{{\bf k},n} \hbar G^{-1}({\bf k}, i\omega_{n}) a_{{\bf k},n}^{*} a_{{\bf k},n} \\ \nonumber
&+ \frac{T}{2} \sum_{{\bf K},{\bf k},{\bf q},n,m,l} a^{*}_{{\bf K} - {\bf k},n - m} a^{*}_{{\bf k},m}a_{{\bf K} - {\bf q},n - l} a_{{\bf q},l}.
\end{align}
Here
\begin{align}\label{eq:grbog}
\hbar G^{-1}({\bf k}, i\omega_{n}) = i \hbar \omega_{n} - \epsilon_{\gamma}({\bf k}) + \mu_{\gamma} - \hbar\Sigma({\bf k}, i \omega_{n}),
\end{align}
and $\hbar\Sigma({\bf k}, i \omega_{n})$ follows from the retarded self energy by Wick rotation of the real frequency to Matsubara frequencies $i\omega_{n}$. This action describes the same equilibrium physics as coming from the Langevin field equation in Eq. \eqref{eq:equil2}, since after a Wick rotation the equations of motion for the field $a_{{\bf k},n}$ are determined by the average of the Langevin equations. Substituting $a_{{\bf 0},0} \rightarrow a_{{\bf 0},0} + \phi$ and requiring that the terms linear in the fluctuations vanish leads to the equation
\begin{align}\label{eq:GP}
\mu_{\gamma} = \hbar c k_{\gamma} + S({\bf 0},0) + Tn_{0},
\end{align}
where $n_{0}$ is the density of condensed photons. This equation determines the chemical potential of the Bose-Einstein condensate of photons. We obtain $\beta \mu_{\gamma} \simeq 90.9$, and according to Eq. \eqref{eq:pola} we have a corresponding polarization of roughly $0.93$. Therefore, almost all molecules are in the excited state. 
\newline
\indent To determine the collective excitations of the condensate over the ground state we consider the action up to second order in the fluctuations. 
\begin{figure}[h!]
\includegraphics{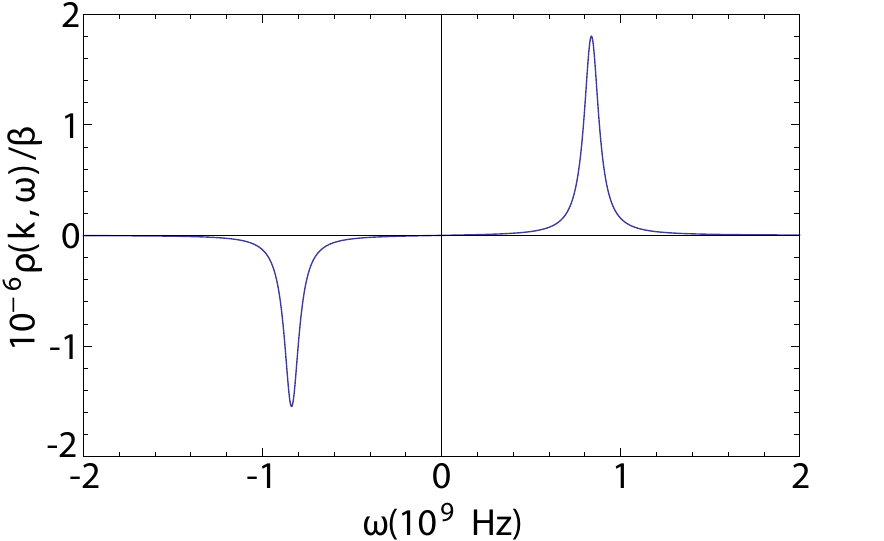}
 \caption{A plot of the spectral function as a function of $\omega$ for $n_{\mathrm{m}} = 9 \cdot 10^{23} \, \mathrm{m}^{-3}$ and $n_{0} = 10^{12} \,\mathrm{m}^{-2}$. In this plot $k \xi \simeq 3.8 \cdot 10^{-2}$.}
 \label{fig:Bogspectral}
\end{figure}
\noindent This is the so-called Bogoliubov approximation. So
\begin{align}
S^{\mathrm{Bog}}[a^{*},a]  = -\frac{1}{2} \sum_{{\bf k},n} u^{\dagger}_{{\bf k},n} \cdot \hbar G^{-1}_{\mathrm{B}}({\bf k},  i\omega_{n}) \cdot u_{{\bf k},n},
\end{align}
\noindent
where
\begin{eqnarray}
u_{{\bf k},n} :=
\left [ \begin{array} {c}
a_{{\bf k},n} \\
a^{*}_{-{\bf k},-n}
\end{array} \right ],
\end{eqnarray}
and
\begin{align}\label{eq:fluc}
- \hbar G^{-1}_{\mathrm{B}}({\bf k}, i\omega_{n}) &=
\left [ \begin{array} {cc}
2 Tn_{0} &Tn_{0} \\
Tn_{0}& 2 Tn_{0}
\end{array} \right ]
 \\ \nonumber &-
\left [ \begin{array} {cc}
\hbar G^{-1}({\bf k}, i\omega_{n}) & 0 \\
0& \hbar G^{-1}({\bf k}, -i\omega_{n})
\end{array} \right ].
\end{align}
Since $\mu_{\gamma}$ is given by Eq. \eqref{eq:GP}, we obtain that $\mathrm{Det}[G^{-1}_{\mathrm{B}}(0,0)] = 0$. Therefore we have a gapless excitation, which agrees with Goldstone's theorem. By Wick rotating and solving for which $\omega$ the determinant of this matrix vanishes, we can determine the dispersions. Since we are interested in the low-energy behaviour, we can use Eq. \eqref{eq:selfap} for the self-energy. In this approximation the dispersions are given by
\begin{align}\label{eq:dispimag}
(1 + \alpha^{2}) \hbar \omega({\bf k}) &= - i \alpha (\tilde{\epsilon}_{\gamma}({\bf k}) + Tn_{0}) \\ \nonumber
&\pm \sqrt{-(\alpha T n_{0})^{2} + \tilde{\epsilon}_{\gamma}({\bf k}) (\tilde{\epsilon}_{\gamma}({\bf k}) + 2 T n_{0})},
\end{align}
with $\tilde{\epsilon}_{\gamma}({\bf k}) = \epsilon_{\gamma}({\bf k}) - \hbar c k_{\gamma}$. The imaginary part of the dispersion relations is always negative and we find a lifetime in the ns regime for $n_{0}$ in the range of $10^{12} - 10^{13} \, \mathrm{m}^{2}$ and excitations for which $k \xi < 0.2$. Recall that k is the norm of the transverse momentum vector ${\bf k} = (k_{\mathrm{x}},k_{\mathrm{y}})$. For excitations with larger momentum the lifetime decreases, until it approaches zero in the limit of $k \rightarrow \infty$. 
\newline
\indent Furthermore, we have the same behavior as was first shown in Ref. \cite{Expol} for a non-equilibrium Bose-Einstein condensate of exciton-polaritons. Also in this case the dispersions become pure imaginary for small momenta. For the numerical values of the experiment and $n_{0} = 10^{12} \, \mathrm{m}^{2}$, we have purely imaginary dispersions for $k \xi  < 2.2 \cdot 10^{-3}$. However, this does not imply that for small momenta there are only decaying quasiparticles at zero energy. This can be seen in the spectral function, which in this case corresponds to the imaginary part of $G_{\mathrm{B};11}({\bf k}, \omega^{+})$. In Figs. \ref{fig:Bogspectral} and \ref{fig:Bogspectral2} we can see the two qualitively different forms of the spectral function. For relatively large momenta, we have two peaks at the real part of the dispersions and the width of the peaks is determined by the imaginary part of the dispersions. In the small-momenta region where both dispersions are purely imaginary, we have a continuum for both negative and positive frequencies. Still, the spectral function has a maximum and a minimum. 
\noindent Therefore, in agreement with the large-momentum case we can also define the position of these extrema as the dispersion. So, contrary to what the analytical dispersion given by Eq. \eqref{eq:dispimag} suggests, also for small momenta the spectral function has a maximum and minimum at non-zero energy.
\newline
\indent Finally, we check if the spectral function satisfies the sum rule given by Eq. \eqref{eq:sumrule}. In the low-frequency approximation for the retarded self energy we can integrate the spectral function analytically, and we obtain
\begin{align}
\int d(\hbar \omega) \, \rho({\bf k},\omega) = \frac{1}{1 + \alpha^{2}}.
\end{align}
Since we are in the Bose-Einstein condensed phase $\alpha \simeq 4.5 \cdot 10^{-3}$, and we satisfy in very good approximation the sum rule. Note that this small deviation from the sum rule is a consequence of making the low-energy approximation to the self energy. Namely, this approximation for the self-energy is only valid for small energies and therefore the high-frequency behaviour is not incorporated correctly. However, without this low-energy approximation the self-energy has both the correct low-energy and high-energy limits, and we indeed find that the spectral function with the full self-energy satsifies the sum rule. 

\section{Conclusion and outlook}
\label{sec:concl}
In this work we constructed a theory for Bose-Einstein condensation of photons in a dye-filled cavity. By using the Schwinger-Keldysh formalism, we obtained a Langevin field equation that describes the complete dynamics of the photons. In particular, it incorporates both the coherent and incoherent dynamics of the gas. Furthermore we found that the finite lifetime of the photons can be captured in a single parameter $\alpha$, that depends on the external laser pumping the dye. In addition, we also found an analytic expression for this parameter. In the homogeneous case we have shown that our theory incorporates the correct equilibrium properties of the gas.
\newline
\indent Subsequently, we calculated the collective modes and spectral functions of the homogeneous photon gas in the normal and Bose-Einstein condensed state. In both phases we found that the lifetime of the photons in the cavity is in the ns regime, which is the same regime as found experimentally in Ref. \cite{BECphoton}. Moreover, we obtained that the dynamics of the collective modes is underdamped. Furthermore in agreement with the results of Ref. \cite{Expol} for exciton-polaritons, we found in the Bose-Einstein condensed phase that dispersions become formally purely imaginary for small momentum. Nevertheless, also for small momentum the spectral function has qualitatively a maximum and minimum at non-zero energy. Finally, in both phases the spectral function is well-behaved and satisfies the sum rule. 
\newline
\indent In future research we will consider in detail the fluctuations, and in particular the phase fluctuations, of the photon Bose-Einstein condensate. For a condensate density of $n_{0} \simeq 10^{12} \,\mathrm{m}^{-2}$, the trap length $l = \sqrt{\hbar/m_{\mathrm{ph}}\Omega} \simeq 7.8 \cdot 10^{-6} \, \mathrm{m}$ is about 2 times smaller than the coherence length $\xi$. However, a condensate density of $n_{0} \simeq 10^{13} \,\mathrm{m}^{-2}$ results in a trap length $l$ that is about 2 times larger than the coherence length $\xi$. Both condensate densities are accessible experimentally \cite{BECphoton}, and therefore we intend to explore both the regime of Bose-Einstein condensation and the quasi-condensate regime.

\acknowledgments
It is a pleasure to thank Dries van Oosten for useful discussions. This work was supported by the Stichting voor Fundamenteel Onderzoek der Materie (FOM), the Netherlands Organization for Scientific Research (NWO), and by the European Research Council (ERC).


\begin{thebibliography}{0}
\expandafter\ifx\csname natexlab\endcsname\relax\def\natexlab#1{#1}\fi
\expandafter\ifx\csname bibnamefont\endcsname\relax
  \def\bibnamefont#1{#1}\fi
\expandafter\ifx\csname bibfnamefont\endcsname\relax
  \def\bibfnamefont#1{#1}\fi
\expandafter\ifx\csname citenamefont\endcsname\relax
  \def\citenamefont#1{#1}\fi
\expandafter\ifx\csname url\endcsname\relax
  \def\url#1{\texttt{#1}}\fi
\expandafter\ifx\csname urlprefix\endcsname\relax\def\urlprefix{URL }\fi
\providecommand{\bibinfo}[2]{#2}
\providecommand{\eprint}[2][]{\url{#2}}

\end{thebibliography}


\begin{thebibliography}{10}
%
\bibitem{Bose}
S. Bose, Z. Phys. {\bf 26}, 178 (1924).
%
\bibitem{Einstein}
A. Einstein, Sitz. ber. Preuss. Akad. Wiss. {\bf 1}, 3-14 (1925).
%
\bibitem{BEC1}
M.H. Anderson, J.R. Ensher, M.R. Matthews, C.E. Wieman, and E.A. Cornell, Science {\bf 269}, 
198–201 (1995).
%
\bibitem{BEC2}
C.C. Bradley, C.A. Sackett, J.J. Tollett, and R.G. Hulet, Phys. Rev. Lett. {\bf 75}, 1687 (1995); 
%
\bibitem{BEC3}
K.B. Davis et al., Phys. Rev. Lett. {\bf 75}, 3969–3973 (1995).
%
\bibitem{BECmagnon}
S.O. Demokritov et al., Nature {\bf 443}, 430 (2006).
%
\bibitem{BECpolariton}
J.Kasprzak et al., Nature {\bf 443}, 409 (2006).
%
\bibitem{BECpolariton2}
R.Balili et al., Science {\bf 316}, 1007 (2007).
%
\bibitem{BECphoton}
J.Klaers, J.Schmitt, F.Verwinger, and M.Weitz, Nature {\bf 468}, 545 (2010).
%
\bibitem{PS}
C.J.Pethick and H.Smith, Bose-Einstein Condensation in Dilute Gases (Cambridge University Press, 2008).
%
\bibitem{BECharm1}
V.Bagnato, D.Kleppner, Phys. Rev. A {\bf 44}, 7439–7441 (1991).
%
\bibitem{BECharm2}
W.J. Mullin, J. Low-Temp. Phys. {\bf 106}, 615–641 (1997).
%
\bibitem{Magnon1}
Sergio M. Rezende, Phys. Rev. B {\bf 79}, 174411 (2009).
%
\bibitem{Magnon2}
Scott A. Bender, Rembert A. Duine, and Yaroslav Tserkovnyak,  Phys. Rev. Lett. {\bf 108}, 246601 (2012).
%
\bibitem{Polariton1}
I. G. Savenko, T. C. H. Liew, and I. A. Shelykh,  Phys. Rev. Lett. {\bf 110}, 127402 (2013).
%
\bibitem{Polariton2}
A. Amo et al., Nature (London) {\bf 457}, 291 (2009).
%
\bibitem{Polariton3}
R. T. Brierley, P. B. Littlewood, and P. R. Eastham,  Phys. Rev. Lett. {\bf 107}, 040401 (2011).
%
\bibitem{Polariton4}
Michiel Wouters, Iacopo Carusotto, and Cristiano Ciuti,  Phys. Rev. B {\bf 77}, 115340 (2008).
%
\bibitem{Polariton5}
D. Read, T. C. H. Liew, Yuri G. Rubo and A. V. Kavokin,  Phys. Rev. B {\bf 80}, 195309 (2009).
%
\bibitem{Theory1}
J.Klaers, J.Schmitt, T.Damm, F.Verwinger, and M.Weitz, Phys. Rev. Lett. {\bf 108}, 160403 (2012).
%
\bibitem{Theory2}
P.Kirton, and J.Keeling, arXiv:1303.3459.
%
\bibitem{Theory3}
W.Fun, M.Yin, and Z.Cheng, arXiv:1301.7136.
%
\bibitem{Theory4}
Yurii Slyusarenko, and Alex Kruchkov, arXiv:1305.1210.
%
\bibitem{Stoofkeldysh}
H.T.C. Stoof, J. Low Temp. Phys {\bf 114}, 11 (1999).
%
\bibitem{Spectrum}
R.R. Birge, Kodak Laser Dyes (Kodak publication JJ-169).
%
\bibitem{Spectrum2}
J.R. Lakowicz, Principles of fluorescence spectroscopy (Springer, 2006).
%
\bibitem{UQF}
H.T.C. Stoof, K.B. Gubbels, D.B.M. Dickerscheid, Ultracold Quantum Fields (Springer, 2009).
%
\bibitem{Expol}
Michiel Wouters and Iacopo Carusotto, Phys. Rev. Lett. {\bf 99}, 140402 (2007).
%
\end{thebibliography}
\end{document}